\documentclass[%
 preprint,
 amsmath,amssymb,
 prd,
]{revtex4-1}

\usepackage{graphicx}
\usepackage{dcolumn}
\usepackage{bm}

\begin{document}

\title{On phenomenological study of the solution of nonlinear GLR-MQ evolution equation beyond leading order }

\author{M. Lalung}
\email{mlalung@tezu.ernet.in}
\author{P. Phukan}
\email{pragyanp@tezu.ernet.in}
\author{J. K. Sarma}
\email{jks@tezu.ernet.in}
\affiliation{
 HEP Laboratory, Department of Physics, Tezpur University, Tezpur 784028, Assam, India }

\date{\today}

\begin{abstract}
We present a phenomenological study of the small-x behaviour of gluon distribution function $G(x,Q^2)$ at next-to-leading order (NLO) and next-to-next-to-leading order(NNLO) in light of the nonlinear Gribov-Ryskin-Levin-Mueller-Qiu (GLR-MQ)evolution equation by keeping the transverse size of the gluons ($\sim 1/Q$) fixed. We consider the NLO and NNLO corrections, of the gluon-gluon splitting function $P_{gg} (z)$ and strong coupling constant $\alpha_s (Q^2)$. We have suggested semi-analytical solutions based on Regge like ansatz of gluon density $G(x,Q^2)$, which are supposed to be valid in the moderate range of photon virtuality$(Q^2)$ and at small Bjorken variable$(x)$. The study of the effects of nonlinearities that arise due to gluon recombination effects at small-x is very interesting, which eventually tames down the unusual growth of gluon densities towards small-x as predicted by the linear DGLAP evolution equation.   
\end{abstract}

\maketitle


\section{Introduction}
\label{intro}
The study of small-x behaviour of gluons is very interesting as the gluons become most abundant partons inside the hadrons and can explain the behaviour of QCD observables like the hadronic cross sections through their initial distributions. Determination of parton distribution functions (PDFs) has always been a fascinating task which has attracted and inspired various collaboration groups like H1, ZEUS collaboration \cite{herapdf2.0}, NNPDF \cite{nnpdf3.0}, CTEQ \cite{cteq} etc. and encouraged many researchers in this field. Moreover, parton densities in hadrons assume key roles in the understanding standard model processes as well as in predictions of such processes at accelerators. But, in the domain of asymptotically small-x, gluons are expected to dominate the proton structure function. Therefore, determination of the gluon density in the small-x region is particularly important. Knowledge of gluon densities or say gluon distribution functions are essential also because of the fact that gluons serve as the basic ingredients in calculation of various high energy hadronic processes, for instance, the mini jet productions or in the computation of inclusive cross sections of hard and collinearly factorizable hadronic collisions. Moreover, in the study of p-p, p-A and A-A processes at small-x, at the relativistic heavy-ion collider (RHIC) \cite{rhic111} and at the CERN's LHC \cite{lhc_prl,lhc111}, the precise knowledge of gluon distribution is essential.   
\par The $x$ and $Q^2$ dependence of the gluon density can be predicted well with much phenomenological success through standard QCD evolution equations. The most basic and widely studied QCD evolution equations at twist-2 level are the Dokshitzer-Gribov-Lipatov-Altarelli-Parisi (DGLAP) \cite{ALTARELLI1977298} and the Ballitsky-Fadin-Kuraev-Lipatov (BFKL) \cite{b,f,kl} equations. The solution of both these equations predict sharp growth of gluon densities at high energies towards small-x, this has been percieved with experimental results of deep inelastic scattering (DIS) experiments at HERA \cite{hera1,hera2,hera3,hera4}. The basic difference between DGLAP and BFKL approach is that the former is based on resummation of large logarithmic $Q^2$ and the later is based on resummation of large logarithmic $1/x$. In recent years DGLAP equation has been established as the standard equation for phenomenological study of DIS experiments as well as for global fits of parton distribution function(PDF) by various groups \cite{dglap1,dglap2,dglap3,dglap4,dglap5} and phenomenological study on the nonlinear modification on DGLAP equations also have been performed \cite{boroun2013approximate,boroun2009shadowing}. 

\par The growth of the quark and gluon densities even though increases abruptly towards large-$Q^2$, but they remain dilute with their transverse size proportional to $1/Q$. At very high energies, region of smaller and smaller values of x can be achieved and the number of gluons increases. These unusual growth of gluons have to be tamed down by means of a mechanism so as to comply with the Froissart bound \cite{froissart} and to preserve unitarity \cite{unitarity}. It is a well known fact that at high energies the hadronic cross section comply with the Froissart bound. The Froissart bound states that the hadronic total cross section cannot grow faster than the logarithm squared of energy, which can be mathematically expressed as $\sigma_{total}= \frac{\pi}{m_{\pi}}(\ln{s})^2$,where s is the square of the centre of mass energy and $m_{\pi}$ is the scale of the strong force. Gluon recombination is believed to serve as the mechanism responsible for a possible saturation of gluon densities at small-x as well as unitarization of the physical cross sections at high energies. The pioneering finding of the geometrical scalling in the description of HERA data \cite{geometrical_hera} and in the production of comprehensive jets in the LHC data \cite{geometrical_lhc} suggests that the phenomenon of saturation occurs in nature \cite{saturate2,saturate222,saturate111}.



\par Although, the DGLAP evolution equations can dileneate the available experimental data in a fairly broad range of $x$ and $Q^2$ with appropriate parametrizations, but while trying to fit the H1 data using DGLAP approach, it fails to provide a good description simultaneously in the region of large-$Q^2$ $(Q^2 > 4\, GeV^2)$ and in the region of small-$Q^2$ $(1.5\, GeV^2 < Q^2 < 4\, GeV^2)$ \cite{hera1,hera2,hera3,hera4}. Also, in the
descripton of the ZEUS data at $Q^2 = 1\,GeV^2$ the DGLAP fit to gluon distribution function can be seen predicticting a negative distribution towards small-x, see this Ref. \cite{zeus_2003}. The gluon recombination effects at small-x introduces nonlinear power corrections to the linear DGLAP equation due to multiple gluon interactions. These nonlinear terms help in taming down of the unusual growth of gluon densities in the kinematics where the QCD coupling constant $\alpha_s$ is still small in the dense partonic system. Gribov, Levin and Ryskin (GLR) \cite{glr} followed by Mueller and Qiu (MQ) \cite{mq1,mq2,mq3} did the first perturbative QCD (pQCD) calculations by considering the fusion of two gluon ladders into one. These calculations, on account of nonlinear corrections in terms of the quadratic term in gluon density, gave rise to a new evolution equation popularly known as the GLR-MQ equation. The GLR equation sums up all the fan diagrams i.e. all the workable 2 $\rightarrow$ 1 ladder combinations which are computed in the double leading logarithmic approximation(DLLA). Later, Mueller and Qiu investigated the contributions of multiparton correlation at the twist-4 approximation based on the Glauber-Mueller model into a further simplified GLR-MQ equation. 

\par In our previous work, we had performed phenomenological study of the gluon distribution function $G(x,Q^2)$ by solving the GLR-MQ evolution equation upto next-to-next-to leading order(NNLO) at small-x\cite{lalung2017nonlinear,phukan2017nnlo}. In that work we studied the gluon distribution function $G(x,Q^2)$ with respect to the resolution scale $Q^2$ at fixed values of the momentum fraction x. However, in this work, we perform a phenomenological study of x evolution of gluon distribution function $G(x,Q^2)$ at fixed value of resolution scale $Q^2$ in light of the GLR-MQ equation in the kinematic range of small-x and moderate $Q^2$. In this kinematic range the gluons are believed to show Regge like behaviour and it is interesting to study the higher order effects on the solution of GLR-MQ equation. Keeping the transverse size ($\sim 1/Q$) of the partons fixed at asymptotically small-x, the gluon density becomes so high that we can practically ignore the gluon contribution coming from the valence quarks $P_{gq}$. The gluon recombination in this region plays vital role. The higher order effects can be incorporated by incorporating the higher order terms of the gluon-gluon splitting function $P_{gg}$ and that of the strong coupling constant $\alpha_s(Q^2)$. We show comparison of our results of gluon distribution function, $G(x,Q^2)$ with that of various collaborations or groups like the CT14 \cite{ct14}, NNPDF3.0 \cite{nnpdf3.0}, PDF4LHC \cite{pdf4lhc}, ABMP16 \cite{abmp16} and MMHT14 \cite{mmht14}. We have also compared our results with the recent HERA PDF data viz. HERAPDF2.0 \cite{herapdf2.0}. We have studied the sensitivity of various parameters on our results and shown a comparison of the nonlinear growth of gluon distribution function as predicted by the GLR-MQ equation with respect to the gluon distribution as predicted by the linear DGLAP equation at asymptotic small-x.

\section{The nonlinear evolution equation}
The GLR-MQ equation is a modified version of the linear DGLAP equation differing from the later by means of an additional term quadratic in gluon density $[xg(x,Q^2)]^2$. This term is due to the correlative interaction between the gluons inside the hadrons. This equation can be depicted as a balance equation where, the net growth of the gluon density $x\Delta g(x,Q^2)$ in a phase cell $\Delta (1/x) \Delta \ln{Q^2}$ is due to the collective effects of both the emission and annihilation processes. This collective effect occurs when the chances for recombination of two gluons into one is as prodigious as the chances for a gluon to split into two gluons. The emission probability of gluons by a vertex $g+g \rightarrow g$ is proportional to $\alpha_s \rho$ and that of the annihilation induced by the same vertex is proportional to $\alpha_s ^2 r^2 \rho ^2$, where $\rho$ = $xg(x,Q^2)/S_{\perp}$ is the density of gluons having the transverse size of $1/Q$ and $S_{\perp} = \pi R^2$ is the target area where the gluons inhabit, R being the correlation radius. At x $\sim$ 1 only the emission of gluons is essential because $\rho << 1$, but in the region of $x \rightarrow 0$, the gluon density $\rho$ grows up and we cannot neglect gluon recombination. In terms of the gluon density $xg(x,Q^2)$, the GLR-MQ equation can be mathematically expressed as \cite{glr_density}
\begin{equation}
\frac{\partial ^2 xg(x,Q^2)}{\partial \ln(1/x) \partial \ln(Q^2)} = \frac{\alpha_s N_c}{\pi}xg(x,Q^2) - \frac{\alpha_s ^2 \gamma }{Q^2 R^2}[xg(x,Q^2)]^2.
\end{equation}
The value of the factor $\gamma$ was calculated to be $\frac{81}{16}$ for $N_c = 3$ by Mueller and Qiu. Now, in the DLLA eq. (1) can be written in terms of $G(x,Q^2)= xg(x,Q^2)$ as

\begin{equation}
\frac{\partial G(x,Q^2)}{\partial \ln Q^2} =  \frac{\partial G(x,Q^2)}{\partial \ln Q^2}\bigg|_{DGLAP} 
-\frac{81}{16}\frac{\alpha_s^2 (Q^2)}{R^2 Q^2}\int_{x}^{1} \frac{dz}{z}G ^2 (\frac{x}{z}, Q^2)
\end{equation}  

eq. (2) is the well known form of the GLR-MQ equation \cite{prytz2001}. Although this equation was not formally derived in this form but, this form of the GLR-MQ equation has been widely studied and applied sucessfully in the phenomenology of small-x QCD and in the description of nonlinear effects by many authors \cite{boroun2009shadowing,DEVEE2014571}. The R.H.S. of this equation consists of two terms, the first term is the linear DGLAP term, while the second term is responsible for shadowing of gluons. The standard DGLAP equation in Mellin convolution space is given by
\begin{equation}
\frac{d}{d \ln Q^2}\begin{pmatrix} f_{q_i} (x,Q^2) \\ f_{g} (x,Q^2) \end{pmatrix} = \Sigma_{j}   \begin{pmatrix}
P_{q_i q_j}(z) & P_{q_i g} (z)  \\ P_{g q_j}(z) & P_{gg} (z) \end{pmatrix}\otimes\begin{pmatrix} f_{q_i} (x,Q^2) \\ f_{g} (x,Q^2) \end{pmatrix},
\end{equation}

where the convolution $\otimes$ reperesents the prescription $f(x)\otimes g(x) = \int_x^1 (dz/z)f(z)g(x/z)$,  $f_{q}$ and $f_{g}$ are the quark and gluon distribution function respectively and $P_{q_i q_j}$, $P_{q_i g}$, $P_{g q_j}$ and $P_{gg}$ are the parton splitting functions \cite{ALTARELLI1977298}. We neglect the contribution coming from the splitting function $P_{g q}$ in the small-x gluon rich region. Also, one thing we can notice from eq. (2) is that the size of the nonlinear term depends on the correlation radius R. When the value of R is comparable to the radius of the hadron ($R_h$), the shadowing corrections are negligibly small, whereas for $R<<R_h$, the shadowing corrections play vital role. We expect, as R grows up the gluon distribution function $G(x,Q^2)$ predicted by eq. (2) will become steeper and steeper. So, the correlation radius here is an important factor which can control the growth of $G(x,Q^2)$.
\par In eq. (2), if we remove the nonlinear term then the equation just yields the linear DGLAP evolution of gluon distribution function $\widetilde{G}(x,Q^2)$. The gluon distribution function $G(x,Q^2)$  predicted from eq. (2), is expected to rise slowly with decreasing $x$ for fixed-$Q^2$ in comparison to $\widetilde{G}(x,Q^2)$. We define a parameter $R_T$ such that $R_T = G(x,Q^2)/\widetilde{G}(x,Q^2)$, this parameter quantifies the amount of taming achieved with respect to the linear DGLAP growth of gluon distribution function as $x$ decreases. We go on defining another parameter $R_r$ such that $R_r = {G(x,Q^2)^{R=2}}/{G(x,Q^2)^{R=5}}$. The value of $R = 5$ $GeV^{-1}$ means that the gluons are populated around the size of the proton and $R=2$ $GeV^{-1}$ signifies the gluons concentrated on the hotspots.

\section{\label{sec:citeref}The Regge approximation}

One of the interesting phenomenon observed at HERA was the rise of proton structure function $F_2$ towards small-x corresponding to a rising cross section $\sigma_{{\gamma}^\ast p}$ with increasing invariant mass $W^2$ ($\sim \frac{Q^2}{x} $)of the produced hadronic state. In the framework of Regge theory, this behaviour of structure function plays important role in understanding the behaviour of the observables, which predicts that the total cross section varies $\sigma^{tot} (ab)$ varies as $\sum_i \beta_i s^{\alpha_i -1}$, where $\alpha_i$ is the Regge trajectory and $\beta_i$ are the residue functions. According to Regge theory, at small-x, the behaviour of gluons and sea quarks are controlled by the same singularity factor in the complex plane of angular momentum. Small-x behaviour of the sea quarks and antiquarks as well as the valence quarks distributions are given by the power law $q_{val}(x)\sim x^{-\alpha}$, where the Regge intercept $\alpha =1$ corresponds to a pomeron exchange of the sea quarks and antiquarks while that of the valence quark is given by $\alpha=0.5$. The small-x proton structure function $F_2$ is related to $\sigma^{tot}_{\gamma^{\ast}p}$ which implies $F_2 \sim x^{1-\alpha_i}$. In the Regge inspired model developed by Donnachie-Landshoff(DL)\cite{regge_zeus,cudell1999perturbative,donnachie2002proton}, $\beta_i$ are assumed to be dependent on $Q^2$ and the intercept $\alpha_i$ are independent of $Q^2$. \par  In DL model the HERA data could be fitted very well on adding a hard exchanged pomeron to that of the soft pomeron in the Regge theory. In this way, the addition of a hard pomeron could describe the rise of structure function $F_2$ at small-x. The simplest fit to the small-x data corresponded to $F_2 (x,Q^2)= A(Q^2)x^{-\epsilon_0},$ with $\epsilon_0 = 0.437$ \cite{regge_zeus,cudell1999perturbative,donnachie2002proton}.
\par Thus, in this Regge inspired model, the high energy hard hadronic processes are predominated by a hard pomeron exchange with the intercept of $1+\epsilon_0$, $\epsilon_0$ is the Regge intercept. The mellin transform of $F_2$ would have a pole at $j= 1+ \epsilon_0$, where its origin is perturbative QCD based on summation and resummation of small-x logarithms. The logarithms of x become large in the small-x regime and cannot be neglected, they need to be resummed based on the BFKL equation\cite{fadin1975pomeranchuk,kl}. Resummation of these small-x contributions is performed in accordance with the so-called $k_T$ factorization scheme. Solution to the leading order BFKL equation leads to a pole of $j= 1+ 4N_c\alpha_s \ln 2 /\pi$ in the anglular momentum plane corresponding to a hard pomeron intercept. 
\par Donnachie and Landshoff in their work also showed that the result of integration of the differential equation $\frac{\partial F_2 (x,Q^2)}{\partial \ln Q^2}\sim P_{qq} \otimes G(x,Q^2)$ at small-x for the gluon distribution function is described by the exchange of a hard pomeron i.e. $G(x,Q^2)= A_G (Q^2)x^{-\epsilon_0}$\cite{regge_zeus,0954-3899-26-5-326,golec,cudell1999perturbative,donnachie2002proton,fadin1975pomeranchuk}.     
\par Towards the small-x region of DIS processes, it is believed to have a greater possibility in exploring the Regge limit of perturbative QCD (pQCD). Models based on Regge ansatz provide frugal paramterizations of the parton distribution functions, $f (x,Q^2) = A(Q^2)x^{-\lambda}$ , where $\lambda$ is pomeron intercept minus one. This type of behaviour of the Regge factorization of the structure function $F_2 ^{c\bar{c}}$ has successful experimental back up in the description of the DIS data of ZEUS in the kinematics of $x < 0.07$ and $Q^2 < 10$ $GeV^2$ \cite{regge_zeus}. \par We, therefore, proceed by considering a simple form of Regge like behaviour given as

\begin{equation}
G(x,Q^2)= \chi (Q^2) x^{-\lambda_G},
\end{equation}

\noindent where $\lambda_G = (4N_c\alpha_s  \ln{2})/\pi$, as appropriate for $Q^2 > 4$ $GeV^2$ \cite{lambda_g} and $N_c$ is the number of color charges. The value of $\lambda_G$ thus depends on the choice of $\alpha_s$. This implies that $G^2(x/z,Q^2)= G^2 (x,Q^2) z^{2\lambda_G}$. So, the Regge intercept $\lambda_G$ will play a central role in our calculations. This type of form is believed to be valid in the region of small-x and intermediate range of $Q^2$, where $Q^2$ must be small but not so small that $\alpha_s (Q^2)$ is too large. But, it is to note that the Regge factorization cannot be a good ansatz in the entire kinematic region. Regge theory is  apparent to be applicable when the invariant mass $W^2(= (1/x - 1)/Q^2)$ is much greater than all other variables. So, the kinematic range which in fact we are considering viz., $10^{-5} \leq x \leq 10^{-2}$ and $5\, GeV^2 \leq Q^2 \leq 30\, GeV^2$ fall in the Regge regime.

\section{Semi-analytical solution beyond leading order}
We incorporate the higher order terms of the splitting function $P_{gg}$ and the QCD coupling constant $\alpha_s (Q^2)$. Both of these terms can be expanded perturbatively to include higher order contributions coming from higher twist effects. Considering the next-to-leading order (NLO) and next-to-next-to-leading order (NNLO) terms, $\alpha_s$ can be written as 
\begin{equation}
	\alpha_s ({Q'}^2)^{NLO} =  \frac{4\pi}{\beta_0 \ln{{Q'}^2}}(1 - b \frac{\ln{(\ln{{Q'}^2})}}{\ln{{Q'}^2}}),
\end{equation}
\begin{equation}
	\alpha_s ({Q'}^2)^{NNLO} =  \frac{4\pi}{\beta_0 (\ln{{Q'}^2})^2} \bigg\lbrace \ln{{Q'}^2} - b \ln{(\ln{{Q'}^2})} 
	b^2\big( {\ln^2{(\ln{{Q'}^2}})} - \ln{(\ln{{Q'}^2})} -1   \big) + c                                \bigg\rbrace,
\end{equation} 
where $b = \frac{\beta_1}{\beta_2^2}$, $c = \frac{\beta_2}{\beta_0^3}$ $\beta_0 = 11- \frac{2}{3}N_f$,  $\beta_1 = 102- \frac{38}{3}N_f$ and $\beta_2 = \frac{2857}{2} - \frac{6673}{28}N_f + \frac{325}{54}N_f ^2.$

Here we take the number of flavors $N_f =4$ and we use the notation ${Q'}^2 = Q^2 / \Lambda ^2$, where $\Lambda$ is the QCD cut off parameter. The splitting function $P_{gg}$ can also be expanded in powers of $\frac{\alpha_s ({Q'}^2) }{2\pi}$ as follows:
\begin{equation}
	P_{gg} (z,{Q'}^2) =  T P_{gg} ^{(0)} (z) + T^2 P_{gg} ^{(1)} (z) + T^3 P_{gg} ^{(3)} (z),
\end{equation}
where $T\equiv T({Q'}^2)= \alpha_s ({Q'}^2)/2\pi$ and $P_{gg} ^{(0)} (z)$, $P_{gg} ^{(1)} (z)$ and  $P_{gg} ^{(0)} (z)$ are the LO, NLO and NNLO terms of the gluon-gluon splitting function respectively \cite{nnlo_splitting}.

\begin{equation}
P_{gg} ^{(0)}(z)= 6\left(\frac{1-z }{z }+\frac{z }{(1-z )_+}+z (1-z )\right)+\left(\frac{11}{2}-\frac{2N_f}{3}\right)\delta
(1-z ).
\end{equation}
The denominator of the second term in the RHS of $P_{gg}^{(0)} (z)$
in written terms of what is known as the `+ prescription'. This indicates the cancellation of singalurity that is appearing at $z=1$ through
\begin{equation}
\int_0^1 \frac{f(z)}{(1-z)_+}dz=\int_0^1 \frac{f(z)-f(1)}{1-z}dz.
\end{equation}

The NLO correction to the gluon-gluon splitting function is given by
\begin{align}
P_{gg}^{(1)} (z)=&C_F T_f \left\{-16+8 z +\frac{20 z ^2}{3}+\frac{4}{3 z }-(6+10 z ) {\ln z}-2 (1+z ) \ln
^2 z \right\} \nonumber  \\
&+N_c T_f \left\{2-2 z +\frac{26}{9} \left(z ^2-\frac{1}{z }\right)-\frac{4}{3} (1+z ) \text{ln$z $}-\frac{20}{9}p(z
)\right\}  \nonumber
\\
&+N_c^2 \bigg\{\frac{27 (1-z )}{2}+\frac{67}{9} \left(z ^2-\frac{1}{z }\right)-\left(\frac{25}{3}-\frac{11 z }{3}+\frac{44 z^2}{3}\right) \text{ln$z $} \nonumber  \\
&+4(1+z ) \ln ^2 z +\left(\frac{67}{9}+\ln ^2 z -\frac{\pi ^2}{3}\right) p(z ) \nonumber   \\
&-4\text{ln$z$} \ln  (1-z ) p(z )+2 p(-z ) S_2(z )\bigg\},
\end{align} where 
$p(z )=\frac{1}{1-z }+\frac{1}{z }-2+z (1-z )$  and
$S_2(z )=\int _{\frac{z }{1+z}}^{\frac{1}{1+z }}\frac{\text{dz}}{z} \ln \left(\frac{1-z}{z}\right) \underset{z }{\overset{\text{small}}{\longrightarrow
} }\frac{1}{2}\ln ^2z -\frac{\pi ^2}{6}+O(z )$

Finally, the NNLO corrections to the gluon-gluon splitting function is given by
\begin{align}
P_{gg} ^{(2)} (z)=&2643.52 D_0+4425.89 \delta  (1-z )+3589 L_1-20852+3968 z -3363 z ^2      \nonumber  \\
&+4848 z ^3+L_0 L_1 \left(7305+8757 L_0\right)+274.4
L_0-7471 L_0^2+72 L_0^3-144 L_0^4  \nonumber  \\ &+\frac{14214}{z }+\frac{2675.8 L_0}{z }+N_f \bigg\{-412.172 D_0-528.723 \delta  (1-z )-320 L_1  \nonumber        \\ &-350.2\,
+755.7 z -713.8 z ^2+559.3 z ^3+L_0 L_1 \left(26.15\, -808.7 L_0\right)+1541 L_0 \nonumber            \\ &+491.3 L_0^2+\frac{832 L_0^3}{9}+\frac{512 L_0^4}{27}+\frac{182.96}{z
}+\frac{157.27 L_0}{z }\bigg\}   \nonumber       \\
&+N_f^2 \bigg\{-\frac{16 D_0}{9}+6.463 \delta  (1-z )-13.878\, +153.4 z -187.7 z ^2  \nonumber      \\ &+52.75 z
^3-L_0 L_1 \left(115.6\, -85.25 z +63.23 L_0\right)-3.422 L_0      \nonumber    \\ &+9.68 L_0^2-\frac{32 L_0^3}{27}-\frac{680}{243 z }\bigg\},
\end{align}
where $D_0=\frac{1}{(1-z)_+}, L_0=\ln z$, $L_1= \ln (1-z )$,
$C_F=\frac{N_c^2-1}{2N_c}\, \text{and}\, T_f=\frac{1}{2}N_f$.\\

\par In terms of the variable ${Q'}^2$, eq. (2) can be re-wriiten in the form
\begin{equation}
\frac{\partial G(x,{Q'}^2)}{\partial ln {Q'}^2} =  \frac{\partial G(x,{Q'}^2)}{\partial \ln {Q'}^2}\bigg|_{DGLAP} -\frac{81}{16}\frac{\alpha_s^2 ({Q'}^2)}{R^2 {Q'}^2 \Lambda ^2}\int_{x}^{1} \frac{dz}{z}G ^2 (\frac{x}{z}, {Q'}^2).
\end{equation}   
So, we notice that choice of the QCD cut off parameter $\Lambda$ is also important in this equation. The ${Q'}^2$ dependence of $\alpha_s$ makes the nonlinear eq. (12) more complicated to solve at NLO and NNLO. Thus, we define two new parameters $T_0$ and $T_1$ such that $T^2 \approx T\cdot T_0$ and $T^3 \approx T\cdot T_1$ respectively. These two parameters are estimated using nonlinear model fiiting techniques in the region of 5 $GeV^2 \leq Q^2 \leq$ 30 $GeV^2$, which is the region of our interest in this work. These parametrizations simplify the nonlinear equation which then can be solved for $x$. The paramater statitics of the fitted model are listed in table I.

\begin{figure}[t]
		\centering
\includegraphics[width=0.45\textwidth]{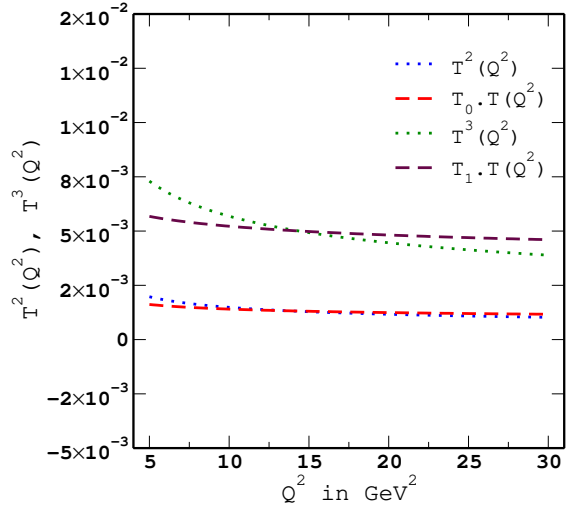}
\caption{\label{fig:fitting_parameter} Parameter fitting of $T^2 (Q^2)$ and $T^3 (Q^2)$ with respect to $Q^2$. }
\end{figure}

\begin{table}[h]
\centering
	\caption{\label{tab:table1}%
		Table showing the parameter statistics of the parameters $T_0$ and $T_1$ 
	}
	
		\begin{tabular}{ccccl}
			\hline\noalign{\smallskip}
			\textrm{\bf Parameter}&
			\textrm{\bf Estimate}&
			\textrm{\bf Standard Error}&
			\textrm{\bf t-Statistic}&
			\textrm{\bf P-Value}\\
			\hline\noalign{\smallskip}
			$\bf T_0$  & 0.0364162 & 3.41393$\times 10^{-4}$ & 106.669 & 8.104$\times 10^{-105}$\\
			$\bf T_1$  & 0.00135821 & 0.164091$\times 10^{-4}$ & 82.7717 & 6.32517$\times 10^{-94}$\\ 
			\hline
		\end{tabular}
\end{table}

\subsubsection{On considering upto NLO terms}
Now, in order to obtain the solution of eq. (12) by considering upto the NLO terms of the running coupling constant and the gluon-gluon splitting function, we put the corresponding terms in eq. (12). After few algebra and simplifications, the GLR-MQ equation in terms of the variable ${Q'}^2$ takes up the form of following partial differential equation
\begin{equation}
\frac{\ln({Q'}^2)}{\left[1-b \ln(\ln({Q'}^2) )/{\ln({Q'}^2})\right]} \frac{\partial G(x,{Q'}^2)}{\partial \ln({Q'}^2)}
=\psi (x)G(x,{Q'}^2) - \phi(x)  \frac{G^2 (x,{Q'}^2)}{{Q'}^2 \Lambda ^2}.
\end{equation}
\par The nonlinearity in  eq. (13) is definitely seen in terms of the quadratic term in $G(x,Q^2)$, in addition to this, rather complicated functional form of the running coupling constant in ${Q'}^2$ makes it difficult to have an exact analytical solution. However, the solution to eq. (13) has the following functional form 

\begin{equation}
G(x,{Q'}^2)= \frac{e^{\frac{b \psi (x)}{\ln({Q'}^2)}} \ln({Q'}^2)^{(1+\frac{b}{\ln({Q'}^2)})\psi (x)}}{C + \int_{1}^{\ln({Q'}^2)} \frac{e^{\zeta (x,y)} \phi (x) (y-b \ln y)dy}{y^2}}.
\end{equation}

\noindent Here C is a constant of integration which can be determined using suitable initial condition of gluon distribution function for a fixed-${Q'}^2$ at a given $x_0 (> x)$. We take input value of gluon distribution function $G(x_0,{Q'}^2)$ from PDF4LHC15 PDF data at a larger value of $x$ ($x_0$) for a given value of ${Q'}^2$. We have taken the input value at momentum fraction $x=10^{-2}$ from PDF4LHC15, this set is based on the 2015 recommendations \cite{pdf4lhc} of the PDF4LHC working group. PDF4LHC15 PDFs contain combinations of more recent CT14 \cite{ct14},
MMHT2014 \cite{mmht14}, and NNPDF3.0 \cite{nnpdf3.0} PDF ensembles and are  based  on  an  underlying  Monte  Carlo  combination  of these three PDF groups, denoted  by  MC900. Thus, the NLO x-evolution of $G(x,{Q'}^2)$ for smaller-x($x<x_0$) at fixed-${Q'}^2$ with proper initial condition is given by
\begin{equation}
G(x,{Q'}^2)= \frac{G(x_0,{Q'}^2) e^{\frac{b \psi (x)}{\ln({Q'}^2)}} \ln({Q'}^2)^{(1+\frac{b}{\ln({Q'}^2)})\psi (x)}}{{\ln({Q'}^2)}^{(1+\frac{b}{\ln({Q'}^2)})\psi (x_0)} e^{\frac{b \psi (x_0)}{\ln({Q'}^2)}}+G(x_0,{Q'}^2) \int_{1}^{\ln({Q'}^2)} \frac{\lbrace e^{\zeta (x,y)}\cdot \phi (x)- e^{\zeta (x_0,y)} \phi (x_0)\rbrace(y-b \ln y) dy}{y^2}}  ; 
\end{equation}
 $\forall\, x \leq x_0$, where the functions involved are given by
\begin{equation*}
\begin{split}
\small
\psi(x)=& \frac{12}{\beta_0}\bigg\lbrace \frac{11}{12}-\frac{N_f}{18}+\ln(1-x)+\frac{2}{2+\lambda_G} -\frac{2x^{\lambda_G +2}}{\lambda_G +2}+\frac{x^{\lambda_G}}{\lambda_G}-\frac{1}{\lambda_G}-x+1\bigg\rbrace+ \frac{2T_0}{\beta_0}\int_x^1 dz P_{gg}^{(1)}(z)z^{\lambda_G}, \\
\zeta(x,y)&= \frac{b \psi(x)}{y}-y +  \psi(x)\ln{y} + b\psi(x)\frac{\ln{y}}{y}, \phi (x)= T_0\cdot\frac{81\pi^2}{2\beta_0 R^2 \Lambda^2}\bigg( \frac{1 - x^{2\lambda_G}}{2\lambda_G} \bigg)
\end{split}
\end{equation*}

\subsubsection{On considering upto NNLO terms}
Now considering upto NNLO terms the GLR-MQ equation upto NNLO takes up the following form:
\begin{equation}
\small
\frac{(\ln{{Q'}^2})^2}{\left[\ln{{Q'}^2}-b \ln(\ln{{Q'}^2}) - b^2 \ln(\ln{{Q'}^2}) + b^2 \ln((\ln{{Q'}^2})^2) - b^2 +c \right]} \frac{\partial G(x,{Q'}^2)}{\partial \ln({Q'}^2)}
=\gamma (x)G(x,{Q'}^2) - \phi(x) \frac{G^2 (x,{Q'}^2)}{{Q'}^2 \Lambda ^2}
\end{equation}
We follow the same procedure to solve this partial differential equation as in the NLO case. Finally, after putting the initial conditions the x-evolution solution (for $x \leq x_0$) of this equation for fixed-${Q'}^2$ is given by
\begin{equation}
\begin{split}
\small
&G(x,{Q'}^2)=\\ &\frac{  G(x_0, {Q'}^2)e^{(\frac{b}{\ln{Q'}^2} - \frac{c}{\ln{Q'}^2} - \frac{b^2 \ln ^2{(\ln{Q'}^2)}}{\ln{Q'}^2}  )\gamma (x)} (\ln{Q'}^2)^{(1 + \frac{b}{\ln{Q'}^2}- \frac{b^2}{\ln{Q'}^2})\gamma (x)}                    }{     e^{(\frac{b}{\ln{Q'}^2} - \frac{c}{\ln{Q'}^2} - \frac{b^2 \ln ^2{(\ln{Q'}^2)}}{\ln{Q'}^2}  )\gamma (x_0)} (\ln{Q'}^2)^{(1 + \frac{b}{\ln{Q'}^2}- \frac{b^2}{\ln{Q'}^2})\gamma (x_0)} + G(x_0,{Q'}^2) {\int_{1}^{\ln{Q'}^2} \frac{ \left(\phi(x)e^{\Delta(x,y)} - \phi(x_0)e^{\Delta(x_0,y)} \right)\eta(y)      }{y^2}dy}          }, \\
&\Delta(x,y)= \bigg( \frac{b}{y}-\frac{c}{y} + \ln{y} + \frac{b\ln{y}}{y} -\frac{b^2\ln{y}}{y} - \frac{b^2\ln^2{y}}{y}  \bigg)\gamma(x)-y,\gamma(x)= \psi(x) + \frac{2T_1}{\beta_0}\int_x^1 dz P_{gg}^{(2)} (z) z^{\lambda_G},\\
&\eta(y)= - b^2 +c +y -b\ln{y} - b^2\ln{y} + b^2\ln^2{y}.
\end{split}
\end{equation}

After computing all these solutions in terms of the variable ${Q'}^2$, we can return back to our original variable $Q^2$ by just substituting $Q^2/\Lambda^2$ in place of ${Q'}^2$. Thus x-evolution of gluon distribution function $G(x,Q^2)$ from the nonlinear GLR-MQ equation beyond the leading orders can be obtained based on the Regge behaviour of gluons at small-x and moderate-$Q^2$.

\begin{figure}[t]
	\centering
	
	\includegraphics[width=.4\textwidth]{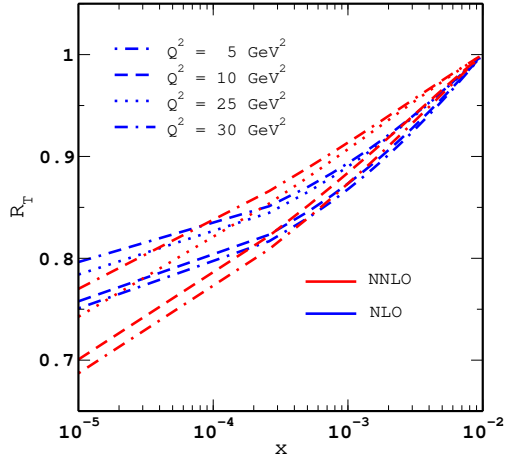} 
	\caption{$R_T$ ratio of the solution of nonlinear GLR-MQ equation to the linear DGLAP equation. }
\end{figure}

\section{Discussions}

So, we have suggested semi-analytical solutions of the GLR-MQ equation at NLO and NNLO based on the Regge like behavior of gluons in the kinematic range of $10^{-5} \leq x \leq 10^{-2}$ and $5 \,GeV^2 \leq Q^2 \leq 30\, GeV^2$. Our solution predicts the x-evolution of gluon distribution function $G(x,Q^2)$ at NLO and NNLO for fixed-$Q^2$ which is also consistent with our previous result at the LO \cite{DEVEE2014571}.  

\par Fig. 1 shows a comparison of $T^2 (Q^2)$ with $T(Q^2).T_0$ and $T^3(Q^2)$ with $T(Q^2).T_1$ for $5 \, GeV^2 \leq Q^2 \leq 30 \, GeV^2$. We have multiplied both $T^3(Q^2)$ and $T(Q^2).T_1$ by a factor of 10, so as to represent all these variations in a single figure. We have determined the value $T_0 = 0.0364162$ and $T_1 = 0.00135821$ for the best fit of the data in the range of our consideration. Table I shows the parameter statistics associated with the fitted parameters. The standard error for the parameter $T_0$ is of the order of $10^{-4}$, while that of the parameter $T_1$ is of the order of $10^{-5}$, in the given $Q^2$ range. From the figure it is also visible that the reduced functions $T_0.T$ and $T_1 . T$ show almost similar distribution with their counterparts $T^2$ and $T^3$  respectively. However, it is to mention that this reduction technique is valid only in the given range of $Q^2$ i.e. from $5\, GeV^2$ to $30\, GeV^2$. In order to study the effect of the standard error of these parameters on our result of gluon distrubution function, we also computed the standard error on $G(x,Q^2)$ arising from the standard error of $T_0$ and $T_1$. In the Fig. 3, the standard error of $G(x,Q^2)$ is expressed in terms of error bar. We can see from the figure that the error bars are very short meaning the effect of these parameters is significantly less.

\par In Fig. 2, we plot the ratio $R_T$ of gluon distribution function $G(x,Q^2)$ predicted from GLR-MQ equation to the gluon distribution function $\widetilde{G}(x,Q^2)$ predicted from DGLAP equation. We have shown a comparison of $R_T$ values for four fixed values of $Q^2$ viz., 5, 10, 25 and 30 $GeV^2$ respectively. We observe that as we go towards small-x, the $R_T$ value decreases, i.e., the taming is more towards small-x for a fixed-$Q^2$. We also observe that on increasing $Q^2$, the $R_T$ value also increases, this means that taming is lesser for higher-$Q^2$ than for low-$Q^2$. This makes sense because the transverse size of the gluons grows up as $1/Q$, smaller the size of gluons lesser is the amount of shadowing. We also observe that the taming of our NNLO solution is more as compared to the NLO solution as x decreases.

\begin{figure}[t]
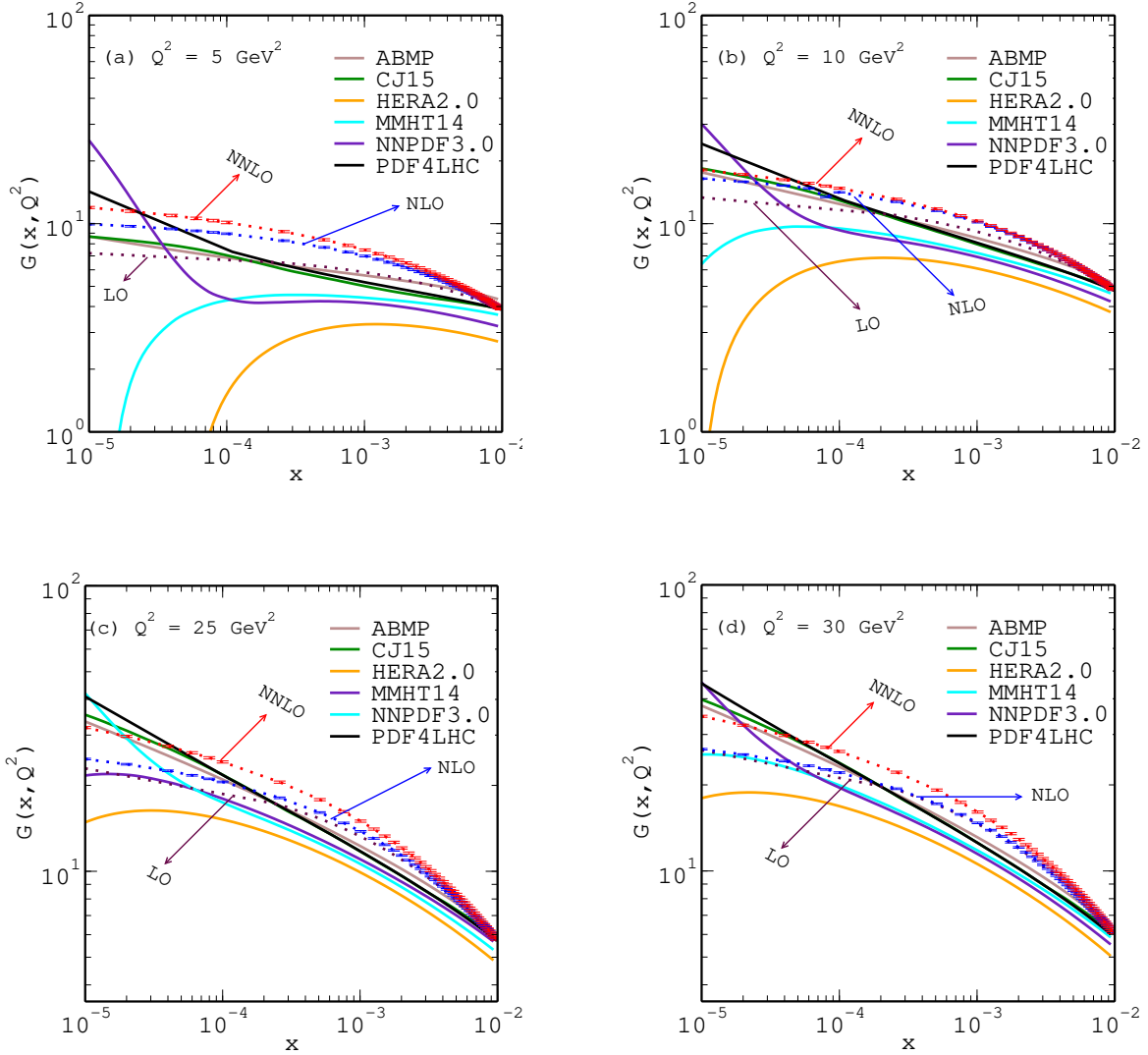

  \label{ fig7} 
  \begin{minipage}[b]{0.45\textwidth}
 
    \includegraphics[width=.95\linewidth]{Q5plot.eps} 

    \vspace{7ex}
    
  \end{minipage}
  \hspace{5ex}
  \begin{minipage}[b]{0.45\textwidth}
    \includegraphics[width=.95\linewidth]{Q10plot.eps} 
    \vspace{7ex}
  \end{minipage}
  \begin{minipage}[b]{0.45\textwidth}
  
    \includegraphics[width=.95\linewidth]{Q25plot.eps} 
    \vspace{7ex}
  \end{minipage}
   \hspace{5ex}
  \begin{minipage}[b]{0.45\textwidth}
 
    \includegraphics[width=.95\linewidth]{Q30plot.eps} 
    \vspace{7ex}
  \end{minipage} 
  
	\caption{$x$-evolution of $G(x,Q^2)$ for R= 2 $GeV^{-1}$ at four different values of $Q^2$ viz. $Q^2$ = 5, 10, 25 and 30 $GeV^2$ respectively. Figure showing a comparison of our results with that of the global fits by various groups. Here, our LO, NLO and NNLO results are represented by dotted violet lines, dotted blue lines and dotted red lines respectively.}
\end{figure}

	

\begin{table*}[t]
\caption{\label{tab:table1}%
Table showing the various parameters for the best fit of the graphs in Fig. 2(a-d) }
\begin{tabular}{ccccccccccccc}
\hline
 &\multicolumn{3}{c}{$\bf Q^2=5$ \bf ($\bf GeV^2$)} &\multicolumn{3}{c}{$\bf Q^2=10$ \bf ($\bf GeV^2$)} &\multicolumn{3}{c}{$\bf Q^2=25$ \bf ($\bf GeV^2$)} &\multicolumn{3}{c}{$\bf Q^2=30$ \bf ($\bf GeV^2$)}\\
 \hline
 {\bf Parameters} &{\bf LO} &{\bf NLO} &{\bf NNLO} &{\bf LO}
&{\bf NLO} &{\bf NNLO} &{\bf LO}
&{\bf NLO} &{\bf NNLO} &{\bf LO} &{\bf NLO} &{\bf NNLO} \\ 
\hline
 {\bf R ($\bf GeV^{-1}$)}&$2$&$2$ &$2$&$2$ &2
&2 &2 &2 &2
&2 &2 &2\\
{\bf $\bf \lambda_G$}&0.36
 &0.35&0.39&0.33&0.32
&0.38&0.325 &0.31 &0.36
&0.32 &0.31 &0.36\\
{\bf $\bf \alpha_s$}&0.136&$0.132$
 &$0.147$&0.125
&0.121&0.143&0.123 &0.117 &0.136
&0.120 &0.117 &0.136\\
{\bf $\bf \Lambda$ (GeV)}&0.3&0.3&0.3&0.3
&0.3&0.3&0.3 &0.3 &0.3
&0.3 &0.3 &0.3\\
\hline
\end{tabular}
\end{table*}

\par In Fig. 3(a-d), we plot our NLO as well as NNLO solutions for $G(x,Q^2)$ in the kinematics of $10^{-5} \leq x \leq 10^{-2}$ at $Q^2 = 5,\,10,\,25$ and $30$ $GeV^2$ respectively. It can be observed that our NNLO result lies slightly above the NLO result. This is due to the additional higher order gluon-gluon splitting terms present in the splitting function $P_{gg} (z)$. The value of the Regge intercept $\lambda_G$ is crucial in this phenomenological study. The strong coupling constant $\alpha_s$ also enters into the picture through the relation $\lambda_G = (4\alpha_s N_c/\pi)\ln{2}$ \cite{lambda_g} which can control the growth of $G(x,Q^2)$. We have shown comparison of our results with those obtained by global DGLAP fits by various collaborations like CT14 \cite{ct14}, NNPDF3.0 \cite{nnpdf3.0}, HERAPDF2.0 \cite{herapdf2.0}, PDF4LHC \cite{pdf4lhc}, ABMP16 \cite{abmp16} and MMHT14 \cite{mmht14}. We have used APFEL tool \cite{apfel} to generate the gluon distribution functions of these collaborations in the kinematics of $10^{-5} \leq x \leq 10^{-2}$ for $Q^2 =5,\, 10,\,25$ and $30$ $GeV^2$. We have used the LHAPDF6 \cite{lhapdf} PDF grids to generate these data. For the best fit of results shown in Fig. 3(a-d), all the parameters that we have considered are listed in the Table II. From this figure, we have seen that while, our results are compatible and close to various groups parametrizations, our results differ significantly from the HERAPDFs. This is beacuse we have taken the input value from the PDF4LHC at $x= 0.01$, and the PDF4LHC itself seems to differ from HERAPDF as can be seen in the figure.

\par In Fig. 4(a-b), we check the sensitivity of R and $\lambda_G$ on our results. In Fig. 4(a), we plot the NLO and NNLO gluon distribution functions $G(x,Q^2)$ in the same kinematic range that we are considering, for four different values of $\lambda_G$ viz., 0.3, 0.4, 0.5 and 0.6 respectively. For reference we take $Q^2 = 30\, GeV^2$ and R = 2 $GeV^{-1}$ respectively. We observe a sharp rise of $G(x,Q^2)$ towards small-x as we increase the value of $\lambda_G$ . We notice that due to the NNLO corrections, $G(x,Q^2)$ rises faster than that of the gluon distribution function $G(x,Q^2)$ when only the NLO terms were incorporated. In Fig. 4(b)
we plot the ratio ($R_r$) between $G(x,Q^2)$ at $R=2$ $GeV^{-1}$ to $G(x,Q^2)$ at $R=5$ $GeV^{-1}$. It can be observed that the value of $R_r$ decrease as x decreases for a fixed-$Q^2$. This can be attributed to the fact that the taming of $G(x,Q^2)$ is more when the gluons are concentrated at the hotspots(R = 2 $GeV^{-1}$) than when they are spread throughout the size of the proton (R = 5 $GeV^{-1}$). As we increase $Q^2$ the ratio $R_r$ shifts upwards as x decreases, this means the taming will be less when $Q^2$ increases on decreasing x. This is again confirmation of the fact that the size of gluons grows as $1/Q$. Also, we observe that the taming is more in NNLO solution than that of the NLO solution as x decreases for a fixed-$Q^2$. This is due to the additional NNLO term present in the gluon-gluon splitting function $P_{gg}$. The sensitivities of the parameters $\lambda_G$ and $R$ on $G(x,Q^2)$ can also be visualized in a three-dimensional picture as shown in Fig. (5).



\begin{figure}[t]
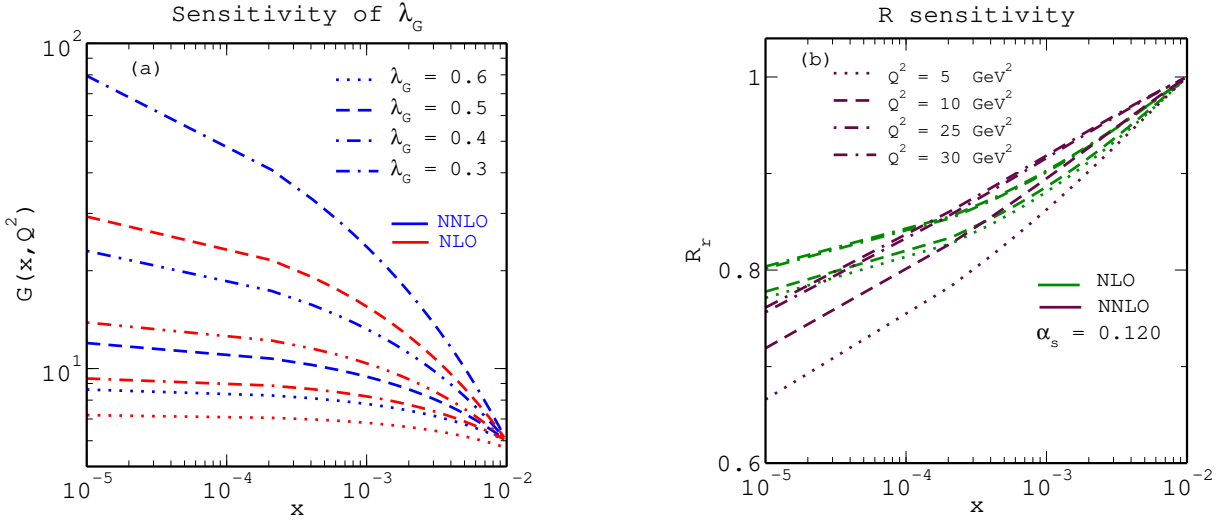

   \label{ fig7} 
   \begin{minipage}[h]{0.45\textwidth}
     \centering
     \includegraphics[width=0.95\linewidth]{lambda.eps} 
 
     
   \end{minipage}
   \hfill
   \begin{minipage}[h]{0.45\textwidth}
     \centering
     \includegraphics[width=0.95\linewidth]{r_sensitivity.eps} 
   \end{minipage}
   \caption{Sensitivity of $\lambda_G$ and R on our results. In Fig. (a) $G(x,Q^2)$ is shown as a function of x at $Q^2$ = 30 $GeV^2$ for four different values of $\lambda_G$ viz., $\lambda_G$ = 0.3, 0.4, 0.5 and 0.6 respectively. In Fig. (b) the variation of $R_r$ is plotted as a function of x for $\alpha_s= 0.120$ for four different values of $Q^2$ viz., $Q^2$ = 5, 10, 25 and 30 respectively.}
 \end{figure}


 \begin{figure}[t]
   \label{ fig7} 
   \begin{minipage}[h]{0.45\textwidth}
     \centering
     \includegraphics[width=0.99\linewidth]{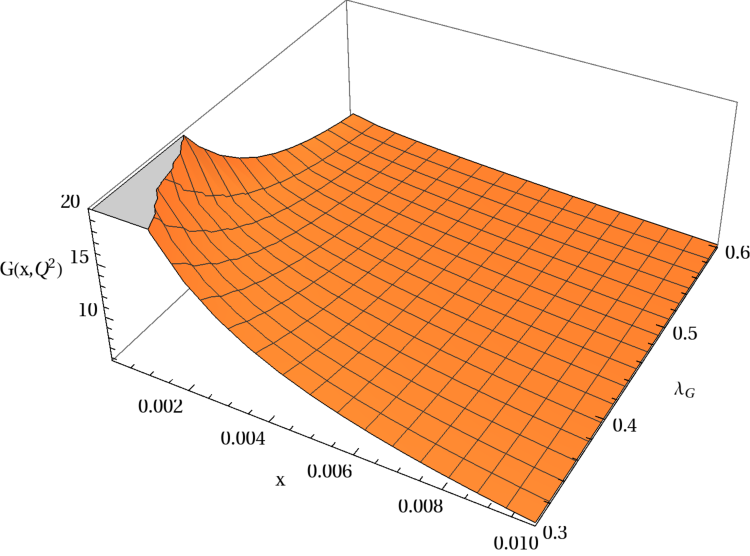} 
 
     
   \end{minipage}
   \hfill
   \begin{minipage}[h]{0.45\textwidth}
     \centering
     \includegraphics[width=0.99\linewidth]{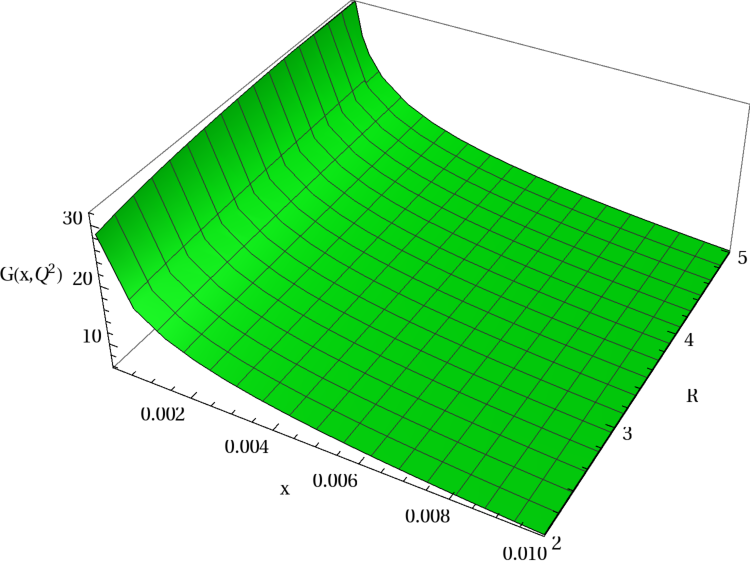} 
   \end{minipage}
   \caption{Sensitivity of $\lambda_G$ and R on our results of gluon distribution function $G(x,Q^2)$ towards small-x (viz. $< 10^{-2}$) at resolution scale $Q^2 = 30\, GeV^2$. On the left figure the sensitivity on gluon distribution function is shown as a function of $x$ and $\lambda_G$ for $R = 2\, GeV^{-1}$ and in the figure on right side, the gluon distribution function is shown as a function of $x$ and $R$ for $\lambda_G=0.35$.   }
 \end{figure}


\section{Summary}
In summary, in this work, we have presented a phenomenological study of the nonlinear effects of  gluon distribution function, $G(x,Q^2)$ in the kinematic range of $10^{-5} \leq x \leq 10^{-2}$ and $5\, GeV^2 \leq Q^2 \leq 30\, GeV^2$ at NLO and NNLO by solving the nonlinear GLR-MQ equation. We have employed the Regge like behavior of gluons in our calculations. We believe that our solutions are valid in the vicinity of saturation scale where it is reasonable to account for the recombination effect to show up because of very high gluon density inside the hadrons and thus our assumptions look natural too. 
The gluon distribution function in our solutions increases as x decreases which is in good agreement with the perturbative QCD fits at small-x. Through our results, we have verified the Regge like behavior of gluons at moderate-$Q^2$ and small-x. It can be observed that our results show almost similar behavior from the results obtained by various global parameterization groups.
\par The advantage of our semi-analytical solution over the exact numerical solutions is that the functional form of the solution makes it easier to perform a phenomenological study of it with respect to various parameters in the solution. For instance, in our phenomenology $\lambda_G$, R, $\alpha_s$ and $\Lambda$ are parameters, in the solutions we can easily control these parameters and we can effectively study the effect of these parameters on our results. In simple words, we can effectively manipulate or control our results by looking upon which parameters to be fixed up, by qualitatively analyzing the form of the solution.  
 \par We have also observed that our solutions are very sensitive to the Regge intercept ($\lambda_G$) and the correlation radius of two interacting gluons ($R$). It can also be observed that compared to our NLO solution, the NNLO solution is more sensitive to $\lambda_G$ and R. In our phenomenological study we also showed that the taming of gluon distribution function becomes more towards small-x at low-$Q^2$ than that of the gluon distribution function towards small-x at a larger-$Q^2$.
 
 So far we have discussed how the saturation phenomenon may show up inside the
 hadrons at very high dense gluon regime at high energy due to nonlinear effects like the gluon
 recombination. Although there has been evidence of saturation from the pioneering finding of geometrical scaling in the description of experimental data at HERA \cite{geometrical_hera} and in
 the production of comprehensive jets in the LHC data \cite{geometrical_lhc}, but conclusive proof of
 saturation is yet to be seen.

 \par
 The Large Hadron electron Collider (LHeC)\cite{0954-3899-39-7-075001}  is a proposed facility which will exploit the new
 world of energy and intensity offered by the LHC for electron-proton scattering,
 through the addition of a new electron accelerator. At LHeC it is expected to reach a wider kinematic
 range of $x$ and $Q^2$ which could not be explored previously. Hopefully, LHeC will give
 a direct evidence of saturation. Apart from probing the saturation regime it is worth
 mentioning that with hundred times the luminosity that was achieved at HERA, some of the
 salient features of the LHeC would be the determination of all light and heavy quark parton
 distributions for the first time, the high precision extraction of the gluon density, the
 determination of the strong coupling constant to per-mil accuracy and the precision study of
 the running of the electroweak mixing angle. LHeC thus will provide a new window on
 QCD and small-x physics and in this domain of small-x physics, we believe that the GLR-MQ equation will be a better candidate than the linear DGLAP equation to explain the nonlinear effects like the shadowing of gluons. Thus, we can conclude that in kinematics, where the density of gluons is very high, the GLR-MQ equation can provide a better understanding of the physical picture than the linear DGLAP equation.

\begin{acknowledgments}
 M. Lalung is grateful to CSIR, New Delhi, India for CSIR Junior Research Fellowship and P. Phukan acknowledges DST, Govt. of India for Inspire fellowship.
\end{acknowledgments}


%



\end{document}